\documentclass[aps,amsmath,amssymb,groupedaddress,twocolumn]{revtex4}
\usepackage{graphicx}

\newcommand{\nc}{\newcommand}
\nc{\bs}{\bigskip}
\nc{\beq}{\begin{equation}}
\nc{\eeq}{\end{equation}}
\nc{\beqa}{\begin{eqnarray}}
\nc{\eeqa}{\end{eqnarray}}

\def\gsim{\mathrel{\rlap{\lower4pt\hbox{\hskip1pt$\sim$}}\raise1pt\hbox{$>$}}}

\begin{document}

\title{What is the entropy of the universe?}  

\author{Paul H.~Frampton}\email{frampton@physics.unc.edu}\affiliation{Department of Physics and Astronomy, UNC-Chapel Hill, NC 27599}
\author{Stephen~D.~H.~Hsu}\email{hsu@uoregon.edu}\affiliation{Institute of Theoretical Science, University of Oregon, Eugene, OR 97403}
\author{Thomas W.~Kephart}\email{tom.kephart@gmail.com}\affiliation{Department of Physics and Astronomy, Vanderbilt University, Nashville, TN 37235}
\author{David Reeb}\email{dreeb@uoregon.edu}\affiliation{Institute of Theoretical Science, University of Oregon, Eugene, OR 97403}

\date{May 2009}

\begin{abstract}
Standard calculations suggest that the entropy of our universe is dominated by black holes, whose entropy is of order their area in Planck units, although they comprise only a tiny fraction of its total energy. Statistical entropy is the logarithm of the number of microstates consistent with the observed macroscopic properties of a system, hence a measure of uncertainty about its precise state. Therefore, assuming unitarity in black hole evaporation, the standard results suggest that the largest uncertainty in the future quantum state of the universe is due to the Hawking radiation from evaporating black holes. However, the entropy of the matter precursors to astrophysical black holes is enormously less than that given by area entropy. If unitarity relates the future radiation states to the black hole precursor states, then the standard results are highly misleading, at least for an observer that can differentiate the individual states of the Hawking radiation.
\end{abstract}

\maketitle

\bigskip
Standard estimates (see Table \ref{tbl:I}) suggest that the entropy of the universe is dominated by that of black holes \cite{entropyofuniverse, Penrose, Kephart:2002bf}. Indeed, a single supermassive black hole, believed to be found in many, if not all, of the $10^{11}$ galactic cores in the visible universe, has more entropy than all the CMB photons combined, according to the conventional area formula. The entropy of the CMB photons in turn dominates that of all other known forms of matter (e.g., stars, planets, galaxies). 

\begin{table}[htb]
\begin{center}
\begin{tabular}{c|c|c}
\hline
{\rm objects} & {\rm entropy} &  {\rm energy} \\
\hline
$10^{22}$ stars  &  $10^{79}$   &   $\Omega_{\rm stars} \sim 10^{-3}$  \\
relic neutrinos  &  $10^{88}$   &   $\Omega_{\rm \nu} \sim 10^{-5}$  \\
stellar heated dust  & $10^{86}$ & $\Omega_{\rm dust} \sim 10^{-3}$ \\
CMB photons  &  $10^{88}$   &   $\Omega_{\rm CMB} \sim 10^{-5}$  \\
relic gravitons  &  $10^{86}$   &   $\Omega_{\rm grav} \sim 10^{-6}$  \\
stellar BHs & $10^{97}$&$\Omega_{\rm SBH} \sim 10^{-5}$\\
single supermassive BH & $10^{91}$ & $10^7 M_\odot$ \\ 
$10^{11} \times 10^7 M_\odot $ SMBH & $10^{102}$ &  $\Omega_{\rm SMBH} \sim 10^{-5}$ \\
holographic upper bound & $10^{123}$ &  $ \Omega = 1$ \\
\hline
\end{tabular}
\end{center}
\caption{Entropies and energies for various systems (using area entropy for black holes). See \cite{TI} for assumptions used in the table.}
\label{tbl:I}
\end{table}

Do black holes dominate the entropy of the universe? If so, what does it mean? In this paper we investigate such questions, under the assumption of unitarity. That is, we {\it assume} that the quantum state of the universe is described by a wavefunction (or a density matrix) which evolves according to a unitary Schr\"odinger-like equation. In particular, we assume that black hole evaporation is unitary, as suggested by the AdS/CFT duality in string theory \cite{AdS}, although of course possibilities alternative to unitarity have not been excluded with certainty \cite{bhreview}. We also note the enormous experimental capabilities required for an observer to have access to the global state of a large system described above (see also \cite{BID}).
 
Statistical entropy $S$ is the logarithm of the number of distinct microstates $\psi$ consistent with the observed macroscopic properties of a system. It is therefore also a measure of uncertainty about the precise quantum state of the system. In fact, the entropy is proportional to the logarithm of the dimensionality of the Hilbert space of $\psi$'s, or, equivalently, to the number of qubits required to exactly specify a particular $\psi$. That is, the information from macroscopic properties (e.g., total energy, size, charge) of the system must be supplemented by $S\,(\ln 2)^{-1}$ qubits in order to fully specify the microscopic state. Statistical entropy, as defined here, happens to equal the von Neumann entropy $-{\rm Tr}\,\rho_e\ln\rho_e$ of the particular mixed state $\rho_e$ that contains all allowed microstates with equal probability. The larger the entropy, the {\it less} is known about the precise quantum state given the macroscopic description. 

We emphasize that the entropy defined above -- the logarithm of the number of possible quantum states consistent with the given macroscopic knowledge about a system at a particular instant in time -- is simply the size (logarithm of dimensionality) of a particular subspace of the Hilbert space of the system. Since this size is preserved by unitary evolution, the entropy we defined does not change in time. It is only coarse grained entropies that increase in time, through loss of information about the state of the system. Our entropy is maximally fine grained. It is, as discussed later in the paper, appropriate for a kind of ``super-observer'' who is unaffected by decoherence and can detect the precise quantum state of the system. 

The objects which contribute most to the entropy of the universe will be those about whose precise quantum state we are most uncertain. The results in Table I indicate that these objects are black holes.

\smallskip

Black holes can dominate the entropy of the universe while only comprising a small fraction of the total energy because their entropy scales as \cite{Hawking, Bekenstein}
\begin{equation}
S_{\rm BH} ~\sim~ A ~\sim~ M^2 
\end{equation}
($A$ is the area and $M$ the mass of the black hole; we suppress prefactors of order one and use Planck units). Note that $S_{{\rm BH}} \sim M^2$ is not coarse-grained entropy, but refers to the number of internal microstates of the black hole, as confirmed by microstate counting in string theory \cite{stromingervafa} or by counting of Hawking radiation microstates \cite{DP}. 

\begin{figure}[ht]
\includegraphics[scale=0.9]{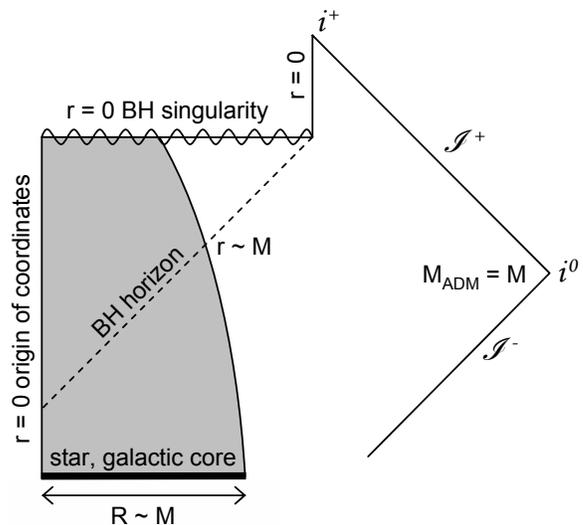}
\caption{Formation of an astrophysical black hole, as happens in our universe, in a conformal diagram: starting from an almost flat spatial hypersurface, a star or galactic core of size $R\sim M$ and containing entropy $<M^{3/2}$ (see (\ref{th1})) collapses to a black hole of mass $M$.}
\label{astrobh}
\end{figure}

However, for non-black hole configurations the following bound \cite{th} holds:
\begin{equation}
\label{th}
S ~<~ A^{3/4}~.
\end{equation}
Note that this bound applies to the microscopic (non-coarse grained) entropy we have defined above; it can be derived as follows. Given a thermal region of radius $R$ and temperature $T$, we have $S \sim T^3 R^3$ and $E \sim T^4 R^3$. Requiring $E < R$ (using the hoop
conjecture -- a criterion for gravitational collapse \cite{hoop,bh}) then implies $T < R^{-1/2}$ and $S <
R^{3/2} \sim A^{3/4}$. The use of a temperature $T$ in this derivation is justified because the entropy of a system of fixed size and total energy is maximized in thermal equilibrium and because of the fundamental connection between statistical mechanics and thermodynamics: the logarithm of the number of microscopic states of a system of fixed size and total energy is, up to corrections that vanish with the volume of the system, the same as the entropy of an equilibrium (canonical) ensemble whose temperature has been adjusted to yield the same average properties. Note, this analysis assumes flat or nearly flat space matter configurations; we will discuss the more general case later. For astrophysical systems on the verge of gravitational collapse (i.e., black hole formation, see Fig.~\ref{astrobh}), $R \sim M$, so the inequality (\ref{th}) becomes
\beq
\label{th1}
S ~<~ A^{3/4} ~\sim~ M^{3/2}~.
\eeq

The bound $S < A^{3/4}$ applies to all known (non-black hole) configurations of matter, including gas clouds, galaxies, ordinary stars and neutron stars. It even applies to our entire cosmological horizon (visible universe) if black holes are excluded, yielding a much stronger upper bound than the holographic one: $S ({\rm universe, ~no~ BH}) < 10^{92}$.

\smallskip

From our discussion so far we can conclude:

\smallskip

${\bf A.}$ A black hole has much more entropy than any ordinary matter configuration of similar size. Equivalently, the Hilbert space required to describe the black hole state is much larger than the Hilbert space describing all matter configurations of similar size.

\smallskip

${\bf B.}$ The Hilbert space describing the {\it subset} of black holes formed from the collapse of ordinary matter configurations whose size is roughly the same as the eventual hole is a \emph{tiny} subspace of the Hilbert space describing all possible black holes of the same mass $M$. This is due to unitary time evolution from the initial matter configuration (having entropy $<M^{3/2}\ll M^2$) to the black hole, during which the size of the Hilbert space is preserved.

\smallskip

Astrophysical black holes are believed to be of the type described in $\bf B$, see Fig.~\ref{astrobh}: all of the mass of the hole comes from the pre-collapse configuration (star, galactic core, etc.); there is a moment (spacelike slice) just before horizon formation when space is still nearly flat and the size of the matter configuration is not very much larger than the eventual hole (i.e., the matter configuration is, say, an order of magnitude larger than the eventual hole), so has size $R\sim M$ and the entropy bound (\ref{th1}) applies.

\begin{figure}[ht]
\includegraphics[scale=0.9]{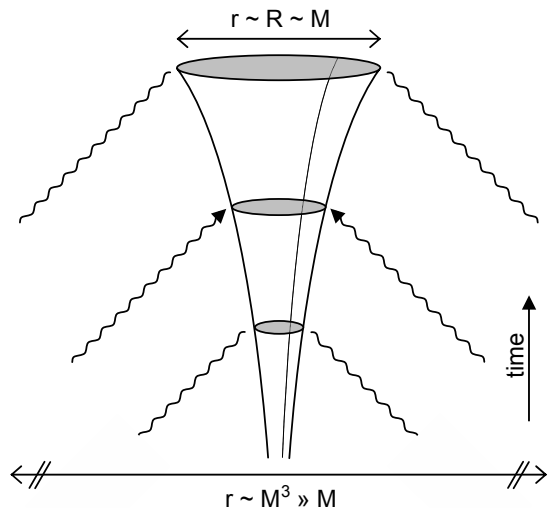}
\caption{Formation of a black hole of mass $M$ by the absorption of thermal radiation similar to Hawking radiation: an initially small black hole continually accretes matter (incoming Hawking radiation), containing total entropy $\sim M^2$ \cite{DP}, and grows to size $R\sim M$. The spatial and timelike extent of the process is $\sim M^3$, much larger than the size and age of the (observable) universe, so none of the existing astrophysical black holes could have possibly formed by this process.}
\label{hawkingbh}
\end{figure}

Our observation concerning astrophysical black holes does not imply that the area entropy of general black holes is spurious \cite{MI}. Consider an alternative process of black hole formation: start with a small hole which then accretes matter, becoming eventually much larger than the original. As a particular example, we might
take the accreting matter to be thermal radiation, as produced in Hawking evaporation, Fig.~\ref{hawkingbh}. In this process a small hole steadily accumulates mass from {\it in}-coming radiation. Since there are at least $\sim \exp{M^2}$ different possible Hawking radiation states \cite{DP}, the resulting hole constructed in this way can be in any of this number of states. That is, the fact that the total entropy of Hawking radiation is $S \sim A$ implies that we {\it can} construct $\exp{A}$ distinct black hole states using this method. However, this process requires large spatial and time extent $\sim M^3$ and does not describe how astrophysical black holes are formed, i.e., by natural processes in the evolution of stars and galaxies \cite{zurekthorne}.

\begin{figure}[ht]
\includegraphics[width=8cm]{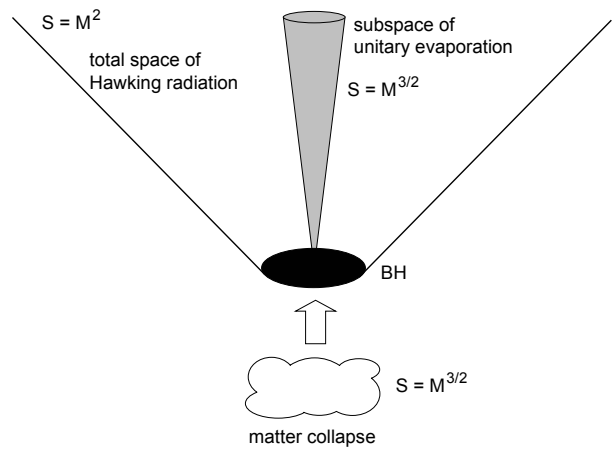}
\caption{Ordinary matter (star, galactic core, etc.) collapses to form an astrophysical black hole. Under unitary evolution, the number of final Hawking radiation states that are actually accessible from this collapse is $\sim \exp M^{3/2}$, i.e.~precisely the number of ordinary astrophysical precursors (\ref{th1}). It is therefore much smaller than the the number of $\sim \exp M^2$ states a black hole, and its eventual Hawking radiation, could possibly occupy if nothing about its formation process were known.}
\label{figure1}
\end{figure}

We arrive at the following conclusion (cf.~also Fig.~\ref{figure1}): 
\smallskip

\noindent {\it If unitarity holds (in black hole formation and evaporation), the future evolution of an astrophysical black hole can produce only a tiny subset ($M^{3/2}$ subspace in Fig.~\ref{figure1}) of all the possible Hawking radiation states indicated by the usual area entropy $M^2$ \cite{cauchyslicefootnote}.} 
\smallskip

As noted, the entropy used in this paper describes the uncertainty in the precise quantum state of a system. If the system is macroscopic the full quantum state is only accessible to a kind of ``super-observer'' who is unaffected by decoherence \cite{decoherence}. Individual observers within the system who have limited experimental capabilities can only detect particular decoherent outcomes. These outcomes arise, e.g., from an effective density matrix that results from tracing over degrees of freedom which are out of the experimenter's control (i.e., which form the ``environment''). In \cite{BID} the experimental capabilities necessary to distinguish decoherent branches of the wavefunction, or, equivalently, to determine the precise quantum state of the Hawking radiation from an evaporated black hole, are discussed. It is shown that a super-observer would either need (at minimum) the capability to make very precise measurements of accuracy $\exp(- M^2 )$ (see also the proposal of Maldacena \cite{eternal} for a specific measurement to determine whether black hole evaporation is unitary), or alternatively the capability to engineer precise non-local operators, which measure a large fraction of the Hawking radiation at once, including quantum correlations (i.e., as opposed to ordinary particle detectors, which only measure Fock state occupation numbers and are neither sensitive to phase information nor to superpositions of different Fock states).

An observer who lacks the capabilities described in the previous paragraph would be unable to
distinguish the states in the $S \sim M^{3/2}$ subspace in Fig.~\ref{figure1} from those in the larger 
$S \sim M^2$ space, assuming the unitarily evaporated states resemble Hawking radiation in gross terms, with large parts of their information hidden in correlations among the emitted quanta or in superpositions of different particle states. In that case, the future uncertainty for ordinary (non-super) observers would be better characterized by the larger $S \sim M^2$ entropy. Putting it another way, ordinary (non-super) observers are {\it forced} by experimental limitations into a coarse grained description of the radiation; they cannot distinguish between most of the radiation states, and for them a {\it different}, coarse grained, entropy is appropriate after evaporation, whose value is $S \sim M^2$. For super-observers, however, the uncertainty in the quantum state does not increase as long as evolution is unitary. From their perspective, a black hole does not have greater entropy than the precursor state from which it formed.

Therefore, for the super-observers described above, the large black hole entropies in Table I do not reflect the actual uncertainties in the (current and future) state of the universe and are in that sense misleading.  A black hole of mass $M$ whose formation history is typical for our universe (e.g., it originated from gravitational collapse of a star or galactic core) satisfies the bound $S<M^{3/2}$ \cite{MI}. Thus, re-evaluating the numbers in Table I, the total entropy of all black holes in our universe is not bigger than the total matter entropy: the dominant uncertainty in the precise state of the universe, at least as far as arises from known physics, is, in fact, due to CMB photons or neutrinos.

As remarked earlier, the $M^{3/2}$ entropy bound (\ref{th1}) was deduced using a flat space approximation (i.e., the relation $V\sim R^3$ between proper volume and spatial extent). One can instead, in a fully general relativistic manner, maximize the entropy of a matter configuration while holding the ADM mass $M$ (the mass seen by a distant observer) fixed. The resulting configurations have been called ``monsters'' \cite{Hsu:2007dr,Sorkin}. Their extremely large entropies are obtained, while holding the local entropy and energy densities fixed, by curving space to create a large internal proper volume without increasing the size of the object as seen by an external observer. The curved space in the construction can also be understood as negative gravitational binding energy which keeps the ADM mass relatively small and fixed although the amount of matter and number of degrees of freedom inside the monster become large. Both monsters and black holes formed from more general processes (i.e., as in Fig.~\ref{hawkingbh} \cite{zurekthorne}) can exceed the $M^{3/2}$ bound and they, indeed, can have area entropy $\sim M^2$. However, monster-like configurations did not (as far as we know) occur in the Universe we inhabit; precursors of astrophysical black holes have entropy $S<M^{3/2}$ and so conform to our argument.

\bigskip

Finally, there is a trivial cosmological argument that implies the black hole entropies in Table I are misleading if unitarity holds. The fine grained entropy we have used throughout this paper (i.e., the logarithm of the number of possible microstates) is conserved under unitary time evolution of the quantum state. Now consider the era of decoupling, when CMB photons decoupled from atoms. In conventional big bang cosmology, there were no black holes in the universe during that epoch. At decoupling, the entropy of thermal photons (and perhaps neutrinos) vastly dominated over all other forms, and consequently the entropy of the CMB modes must still dominate today \cite{note}. Indeed, under ordinary Friedmann-Robertson-Walker expansion (i.e., excluding phase transitions) the total thermodynamical entropy of radiation $S \sim R(t)^3 T(t)^3$ is conserved \cite{KT}. Therefore, most of the uncertainty in the exact quantum state of the universe is due to CMB modes, not to black holes.

\bigskip

\emph{Acknowledgments ---} The authors thank J.~Maldacena and Y.~J.~Ng for useful comments. The authors are supported by the Department of Energy under grants DE-FG02-06ER41418 (PHF), DE-FG02-96ER40969 (SDHH, DR) and DE-FG05-85ER40226 (TWK).
%%%%%%%%%%%%%%%%%%%%%%%%%%%%%%%%%%%%%%%%%%%%%%%%%%%%%%%%%%%%%%%%%
%%%
%%%                     BIBLIOGRAPHY
%%%
%%%%%%%%%%%%%%%%%%%%%%%%%%%%%%%%%%%%%%%%%%%%%%%%%%%%%%%%%%%%%%%%%


\begin{thebibliography}{99}
\bibitem{DP} 
Since the Hawking temperature is $T \sim M^{-1}$, the Hawking radiation has entropy $S \sim \int dQ / T \sim \int dM\,M \sim M^2 \sim A$. This is then, at least for large $A$, also true for the microcanonical definition of entropy which we employ throughout. See also  D.~N.~Page,
  %``Particle Emission Rates From A Black Hole: Massless Particles From An
  %Uncharged, Nonrotating Hole,''
  Phys.\ Rev.\  D {\bf 13}, 198 (1976).
\bibitem{cauchyslicefootnote} In a black hole spacetime, consider two Cauchy slices $\Sigma_1$, preceding black hole formation, and $\Sigma_2$, after complete black hole evaporation, containing all of the Hawking radiation. Unitarity, or one-to-one mapping, requires the dimensionality of the Hilbert spaces of quantum states on $\Sigma_1$ and on $\Sigma_2$ to be equal. Thus, unitarity requires equal microcanonical entropies $S_1$ and $S_2$, which is a puzzle since $S_1 < M^{3/2}$ for ordinary matter but $S_2 \sim M^2$ for the general evaporated state \cite{DP}; this means that, under unitary evolution, only an exponentially small subset of the possible quantum states on $\Sigma_2$ can actually be realized if $\Sigma_1$ only contained ordinary matter states.
\bibitem{note} The number of degrees of freedom of all other systems is tiny compared to that of the CMB (including neutrinos; see Table I without black holes). For this reason only a negligible part of the entropy of the thermal photons and neutrinos could, since the era of decoupling, have been dumped into other forms or into entropy of entanglement with other systems.
\end{thebibliography}
\end{document}